**Ethnicity and gender influence the decision making in a multinational state: The case of Russia.**

Tatiana Kozitsina (Babkina)[a,b*], Anna Mikhaylova[c], Anna Komkova[c], Anastasia Peshkovskaya[d,e], Anna Sedush[a], Olga Menshikova[a,b,f], Mikhail Myagkov[d,g], and Ivan Menshikov[a,b]

[a]Moscow Institute of Physics and Technology (State University), Department of Control and Applied Mathematics. 9 Institutskiy per., Dolgoprudny, Moscow Region, 141701, Russian Federation.

[b]Russian Academy of Science, Federal Research Center "Informatics and Control", Dorodnitsyn Computing Center. 40 Vavilov Str., Moscow, Russian Federation.

[c]North-Eastern Federal University, 58 Belinsky str. Yakutsk, 677000, Russian Federation.

[d]Tomsk State University, Laboratory of Experimental Methods in Cognitive and Social Sciences. 36 Lenin Ave., Tomsk, 634050, Russian Federation.

[e]Mental Health Research Institute, Tomsk National Research Medical Center, Russian Academy of Sciences, 4, Aleutskaya Street, 634014, Tomsk, Russian Federation.

[f]Russian Presidential Academy of National Economy and Public Administration. 82/5, Prospect Vernadskogo, Moscow, 119571, Russian Federation.

[g]Department of Political Science, University of Oregon, 1585 E. 13th Avenue, Eugene, Oregon, 97403, United States.

**\*Corresponding author:** Tatiana Kozitsina (Babkina). Moscow Institute of Physics and Technology (State University), Department of Control and Applied Mathematics. 9 Institutskiy per., Dolgoprudny, Moscow Region, 141701, Russian Federation.

babkinats@yandex.ru

## Abstract

Individuals' behavior in economic decisions depends on such factors as ethnicity, gender, social environment, personal traits. However, the distinctive features of decision making have not been studied properly so far between indigenous populations from different ethnicities in a modern and multinational state like the Russian Federation. Addressing this issue, we conducted a series of experiments between the Russians in Moscow (the capital of Russia) and the Yakuts in Yakutsk (the capital of Russian region with the mostly non-Russian residents). We investigated the effect of socialization on participants' strategies in the Prisoner's Dilemma game, Ultimatum game, and Trust game. At the baseline stage, before socialization, the rates of cooperation, egalitarianism, and trust for the Yakuts are higher than for the Russians in groups composed of unfamiliar people. After socialization, for the Russians all these indicators increase considerably; whereas, for the Yakuts only the rate of cooperation demonstrates a rising trend. The Yakuts are characterized by relatively unchanged indicators regardless of the socialization stage. Furthermore, the Yakutsk females have higher rates of cooperation and trust than the Yakuts males before socialization. After socialization, we observed the alignment in indicators for males and females both for the Russians and for the Yakuts. Hence, we concluded that cultural differences can exist inside one country despite the equal economic, politic, and social conditions.

## Keywords

## 1. Introduction

People from different countries and ethnic groups differ in terms of decision making in economic situations, especially in the context of traditions and habits of citizens (Bond & Forgas, 1984; Roth et al., 1991; Yamagishi, 2003; Yamagishi & Yamagishi, 1994). For example, Spaniards, Africans, and Asians are collective ethnicities compared to English people. Consequently, these ethnicities demonstrate more cooperation in the Prisoner's Dilemma game than English people do (Cox, Lobel, & McLeod, 1991).

With this in mind, a question arises as to whether people from different ethnic groups will behave differently in a multinational state regardless of the fact that ethnicities have lived in the territory of this state since ancient times. On the one hand, social structure and level of education as well as the country's politics and economy are extended to all territories and should minimize ethnicities specific effects. On the other hand, ethnic groups still exist in modern society and thus cannot be neglected.

The answer to the posed question can be found in the Russian Federation because this country is a multinational state that has a vast territory and a diversity of ethnicities populating it. Currently, the Russian Federation includes 85 constituent entities (subjects), which are united in a single nation with similar features, lifestyle, economy, and sociality. Moreover, historically these subjects also unite more than 190 different ethnicities. In fact, many cities have been developing, for a long period of time, being remote and isolated from the rest of a region they belong to, due to long distances and climate patterns, which impact the everyday life and the distinctive features of the development of population.

Do the rates of cooperation, egalitarianism, trust, and reciprocal gratitude for the one ethnic group differ from the other ethnic group in one country? Will the behavior of participants in economic tasks depend not only on ethnicity but also gender? Will the gender patterns revealed for the one ethnicity be found for another one inside one country?

In our attempts to answer these questions, we employed the following two factors: ethnicity and gender. The ethnicities under study were the Russians from Moscow, the city of federal significance and the capital of Russia, and the Yakuts from Yakutsk, the capital of the Sakha Republic (Yakutia). Yakutsk is a remote and thus isolated city populated mostly by the Yakuts, the indigenous nomadic population of eastern type. In 2010, Yakuts represented 49.9% of the population of the Sakha Republic (Yakutia). The population of Moscow consists of 91.6% of ethnic the Russians (1).

Ethnicities and gender differences were investigated using the following three games: Ultimatum Game, Prisoner's Dilemma Game and Trust Game.

The Ultimatum Game characterizes the society from the perspective of social-economic relationships. In the game, it is rational to offer the minimum sum and accept any proposal. However, only even distribution is acceptable in a group. Oosterbeek, Sloof, & Van De Kuilen (2004) conducted meta-analysis of 34 studies related to the Ultimatum Game. It was registered that there was no significant statistical difference in the proposer behavior of people from different cultures (in different countries), however the response to offers varied. According to the studies, participants, on average, offered 40% of the initial sum, but they rejected 16% and lower of the initial sum. On the contrary, Chuah et al. (2007) found that British and Malaysian Chinese participants' proposer behavior patterns were dissimilar (e.g. participants from Malaysia offered more), but there was no statistically significant difference as far as regards the rejection threshold. However the average proposal and rejection thresholds corresponded to the levels registered by Oosterbeek et al. (2004). In the framework of that study, participants were of the same age and had a similar level of education, social and economic background because they studied at the same university. Pairs were formed on a mono- and multicultural basis. A difference in proposer behavior was observed both in mono- and multicultural groups.

Another study using the Ultimatum Game was conducted by Henrich (2000) within the Machiguenga tribe from the Peruvian Amazon. These people's traditional mode of life consisted in living in separate small groups or families without interaction with each other. Despite the fact that now the Machiguenga people are integrated into civilized areas and social life of their country, they still prefer to live in small settlements. Compared to players from industrial countries, Machiguenga players offered about 26% (instead of the minimum 40%), and their rejection threshold was very low and equal to 0.048%. This study confirms the fact that the mode of social life influences social characteristics, viz. egalitarianism.

According to Fedorova et al. (2013),  the Yakuts colonized the Baikal territory, which was also populated by some Mongol tribes (Phillips, 1942). Consequently, individuals belonging to these nations have similar mode of life. Gil-White (2004) conducted experiments using the Ultimatum Game with participants from Mongolia, who led a traditional way of life. The proposal threshold was similar for males (43%) and females (46%). The rejection threshold was near 0%. It can be assumed that these results will correlate with the results obtained from the experiment in Yakutia. This assumption is substantiated with help of findings of a  study performed by Cesarini et al. (2008), who found that the tendency to trust can be inherited genetically. With this in mind, behavior patterns of people from different cultural backgrounds can be similar due to similar genetics.

Tibetan people, who have much in common with the Yakuts in terms of their mode of life, also participated in the Ultimatum Game (Chen & Tang, 2009). Participants from Tibet were compared to ethnic Han Chinese people. The control group consisted of Han Chinese individuals from Singapore. The Tibetans offered less than the Chinese participants, but they were more likely to accept offers regardless of the size of the sum of money, which coincides with the previous results (Gil-White, 2004; Henrich, 2000).

Comparing the above-mentioned three studies, it can be concluded that nations who traditionally lived in small groups are not ready to share; however, they accept any proposals without any doubts. This pattern seems to be predetermined by the climate and life attitudes (i.e. people tend to live in present now rather than envision the future).

The Trust Game is also often used to reveal cultural and gender differences related to trust and reciprocity. As a rule, participants of previously conducted studies (Buchan, Croson, & Dawes, 2002; Johnson & Mislin, 2008) were representatives of different ethnicities; more precisely, they were indigenous people of their countries. This principle enabled us to hypothesize that differences can arise between ethnicities within a multinational country.

Croson & Buchan (1999) proved that there was no correlation between gender and proposals in such countries as the USA, China, Japan, and South Korea, which may have been stipulated by the market price. However, females expressed reciprocity more often than males. In the next part of their study, Buchan et al. (2002) found that cultural differences in trust and reciprocity existed. For example, participants from USA and China were more inclined to trust, whereas participants from South Korea and Japan tended to practice more reciprocity. The effect of group formation on social characteristics also drew the researchers' attention. Participants who formed the groups through the direct-reciprocal exchange demonstrated the highest degree of trust and reciprocity, compared to groups of acquainted individuals, who showed a lower degree of trust and reciprocity, and groups of strangers, whose indicators of trust and reciprocity were lowest. In subsequent studies, Buchan, Croson, & Solnick (2008) presented a refinement of the gender result, according to which only people from the USA had been enrolled in the experiments. Consequently, it was revealed that males made more generous offers, whereas females tended to predominantly return. It is obvious that the economic situation in a country cannot but affect interpersonal relations as far as it concerns trust/mistrust and reciprocity. Johnson & Mislin (2008) ascertained this fact by performing a meta-analysis of 84 experiments in 29 counties. Consequently, it was revealed that culture played an important role in the Trust Game as well as the design of the experiments. In the subsequent study, Johnson & Mislin (2011) complemented the previous findings with the fact that the ultimate result of the game can be affected by the initial in-game sum and the multiplier of the sum transferred to a partner. However, Liebrand & Van Run (1985) proposed a theory, according to which only psychological types have an effect on the behavior of participants' of social dilemmas regardless of their cultural distinctive features. It is supposed that there is a certain correlation between indicators of the Trust Game and the economic situation in the country. It was found that participants from African countries trusted less than participants from North America. Moreover, Balliet & Van Lange, (2013) discovered that cooperation, trust, and the situation in a country are strongly interconnected. In other words, when people are sure that they can rely on their compatriots, it entails an increase in cooperation and trust. It is also important to note that students in China tend to cooperate more than students in the USA including Chinese students in the USA (Hemesath & Pomponio, 1998). Building on this, it can be postulated that in a multinational country with common economy, the welfare factor can be excluded from consideration but cultural factors (viz. mentality, city, and attitude to life) play a role.

Apart from cultural differences, gender makes a significant contribution to participants' behavior, which was proved by numerous studies (Charness & Rustichini, 2011; Eckel & Grossman, 1998; Niall, Mazzoni, & Sbriglia, 2017; Ortmann & Tichy, 1999; Rao et al., 2013). In this context, the principle of dividing participants into pairs in bargaining games is also an important factor. Saad & Gill (2001) registered that in the Ultimatum Game males made more generous offers to females than to males. Females' offers to males and females were virtually at the same level. However, according to Solnick (2001), in the Ultimatum Game males and females tended to offer less if they knew that they were playing with a female. Both males and females chose the highest acceptable offer if the first player (the proposer) had been a female. In fact, it is considered that a female is willing to accept lower sums. In contrast, in the Prisoner's Dilemma Game, other types of pairs demonstrated better cooperation, viz. male-male combinations appeared to be more cooperative than female-female ones (Balliet, Mulder, & Van Lange, 2011). Nevertheless, females were more cooperative in mixed-gender groups. In iterated games, males were more cooperative than females. Females were more cooperative in large groups. In the studies of the effect of group gender on cooperation based on the Public Goods Game, it was found that all-female groups were more cooperative than all-male and mixed-gender ones (Nowell & Tinkler, 1994; Stockard, Van De Kragt, & Dodge, 1988). If participants had a chance to choose a partner, they chose a partner of the opposite sex, and only males made significantly more generous offers to female partners in the Trust Game (Slonim & Guillen, 2010). Similar behavior and gender patterns in the Trust Game were observed in 17 countries (Romano et al., 2017).

Historically, it is considered that individuals from collectivist ethnic groups show the tendency to collaboration and mutual interaction with each other better. Moreover, collective behavior can be shown through the mathematical models (Chartishvili, 2019; Friedkin & Johnsen, 2011; Kozitsin, 2020). In addition to ethnicity, gender also predetermines behavior patterns of individuals in the context of solving social and economic tasks.

In the previous studies (Babkina et al., 2016.; Peshkovskaya et al., 2017; Peshkovskaya, Babkina, & Myagkov, 2018), it was concluded that short-term socialization among participants of the experiments significantly increases the rates of the cooperation even if at the beginning of the experiment, participants were not acquainted with each other. This model of short-term socialization combines the classic social psychology minimal group paradigm with group manipulations that cause a sense of social attachment (Tajfel, 1982; Tajfel & Turner, 1979; Tucker, 1950). It was postulated that socialization is the natural mechanism to promote cooperation and keep it high in group of people (Berkman et al., 2015; Lukinova et al., 2014). In continuation, in this paper, the effect of socialization on individuals from the two different ethnic groups in a multinational state is measured. The goal of the study is to find how socialization depends on ethnicity and gender; how ethnicity and gender are related with each other.

The following research questions are raised and answered:

1. Is there a difference in behavior patterns of the Russians and the Yakuts before and after socialization?
2. Do the Russians and the Yakuts have similar gender-based distinctive features of behavior in economic games, or are these distinctions ethnicity-specific?
3. Is it possible to identify common gender-based and ethnic-based patterns of behavior of the Russians and the Yakuts in the Prisoner's Dilemma Game, Trust Game, and Ultimatum game?

## 2. Materials and Methods

### 2.1. Participants

Participants were recruited by the Moscow Institute of Physics and Technology in Moscow and by the North-Eastern Federal University in Yakutsk. A total of 348 individuals (183 males) participated in 29 experiments. The distribution of participants by ethnicity and gender is presented in Table 1.

**Table 1**

Distribution of the participants by gender and ethnicity.

| Ethnicity | Total number | Number of males | Number of females |
|---|---|---|---|
| the Russians | 216 | 132 | 84 |

| the Yakuts | 132 | 51 | 81 |

The recruitment was performed through posting advertisements on the social networking site *VKontakte* (https://vk.com) and by means of the mobile application *WhatsApp* (https://www.whatsapp.com). For the purpose of each experiment, students, who were unacquainted with each other, were selected. Personal characteristics such as academic major, group, year of study, native language, and ethnic group were considered during the selection. All participants were provided with written and verbal instructions related to the experiment. Experimenters notified participants that all the points that participants will win in the games will be converted to real money (the average win rate was approximately equal to the cost of a full lunch in a cafe). The procedures of the study involving human participants were approved by the Tomsk State University Human Subjects Committee. Written informed consents were obtained from participants. The data associated with this research are available at [link].

### 2.2. Games

The study employed three games: Prisoner's Dilemma Game (PD), Ultimatum Game (UG), and Trust Game (TG).

#### 2.2.1. Prisoner's Dilemma Game

Two individuals participate in each round of the game. They both have two strategies: cooperation or defection. In the standard PD, both participants earn the same profit $R$ in case of mutual cooperation and less profit $P$ in case of mutual defection. If one player cooperates, but another one defects; the cooperator wins the profit $S$, and the defector wins the higher profit $T$. Hence, the following ratio of profits should be maintained: $T > R > P > S$ (Table 2) (Dresher, 1961; Flood, 1958; Rapoport & Chammah, 1965; Tucker, 1980). Defection is more profitable than cooperation regardless of the choice made by the partner-player. Nevertheless, mutual cooperation is more profitable than mutual defection. The Nash equilibrium corresponds to mutual defection (*P,P*), however participants in experiments attempt to extend their interaction to mutual cooperation (*R,R*) (Nowak & Sigmund, 1993).

**Table 2**

Prisoner's Dilemma Payoffs.

| Payoffs | Cooperation | Defection |
|---|---|---|
| Cooperation | R, R | S, T |
| Defection | T, S | P, P |

For the experiments presented in this paper, the following parameters of the PD were set as $R = 5$, $P = 1$, $S = 0$, $T = 10$ ($10 > 5 > 1 > 0$).

### 2.2.2. *Ultimatum Game*

Two individuals play this game for one round. The first player starts the game with 10 points, any part of which he or she proposes to the second player subject. The second player can accept or reject the offer. In case the offer is accepted, both the players win points according to the deal. However, in case the offer is rejected both the players get nothing (Güth, Schmittberger, & Schwarze, 1982). Any offer and any acceptance should be consistent with the Nash equilibrium.

### 2.2.3. *Trust Game*

Two participants play this game for one round. The first player starts the game with 10 points from which she can entrust the second participant with any amount of her initial sum (from 0 to 10 points). The sum proposed to the second participant is tripled. The second participant can return any amount of the tripled sum (Berg, Dickhaut, & McCabe, 1995). In the totally mixed Nash equilibrium, it is not rational to return anything, and therefore it is not rational to trust, which leads to a zero result for both the participants (Cesarini et al., 2008; Delgado, Frank, & Phelps, 2005).

### 2.3. Experimental design

The experimental procedure consisted of three stages.

### 2.3.1. Stage 1. Anonymous playing phase

In order to execute the game, *z-Tree*, a specialized tool for designing and performing experiments in a group of experimental economics, developed at the University of Zurich, was used (Fischbacher, 2007).

Participants played the Prisoner's Dilemma Game and the Ultimatum Game or Trust Game. Eleven game rounds were played. In each round, men and women were randomly divided into pairs and took decisions simultaneously and independently of each other. Each of the 12 men and women could be paired with any other participant of the experiment. In all the rounds, a participant did not know with whom exactly he/she was due to interact. Men and women were notified of the fact that they were playing with one of the 12 participants of the experiment, and each time the partner was changed randomly. After each of the periods, every participant observed on the screen the own result and the anonymous opponent's result.

Points earned at this stage were added to the total win and converted into real money at the end of the game.

### 2.3.2. Stage 2. Socialization phase

At this stage, the participants were involved in social interaction, which consisted in familiarization, communication, and division into groups. The participants memorized each other's names by playing "snowball" (Babkina et al., 2016.). According to the game, first, players were seated in a circle, the first person said his/her name and a personal quality that started with the same letter as the name. Second, the next participant repeated the name along with the quality of the first participant and gave his/her name and quality. Third, the turn was passed over to other participants until it reached the last one, who was due to repeat all the names and personal qualities. Then, in a different order, the participants shared personal information, viz. their hometown, academic major, hobbies, and interests. Following that, two captains were selected from among the participants on a voluntary basis. The captains remained indoors, the other members leave the room, and then, in a random order, they entered the room one by one. Every participant entering the room chose a captain, whose group he/she wanted to join. Consequently, two groups of 6 people were formed. In the end, each group of 6 people was given a task to find 5 common characteristics (i.e. 5 characteristics that united them) and chose a name for their group.

### 2.3.3. Stage 3. Socialized phase

The participants played the Prisoner's Dilemma Game and Ultimatum Game or Trust Game anew. However, unlike at the first stage, the participants interacted only in groups of 6, previously composed during the socialization at the second stage of the experiment. For each round the participants were randomly divided into pairs. Besides, they were informed on the fact that they were interacting with a member of their "own" group, but they did not know who exactly that person was. After each of the periods, every participant observed on the screen the own result and the anonymous opponent's result. Both the games consisted of 15 rounds. The group names, chosen by participants at the socialization stage, appeared on monitors in the Prisoner's Dilemma Game.

Points were added to those obtained at the first stage. As a result, the final prize was generated and could further be converted into a cash reward to be paid to participants.

The schematic representation of the experiment setup included the following parts components:

1. 12 participants are recruited (12 strangers).

2. Participants play 2 games (PD and UG or TG) in a mixed-gender group of 12 people for 11 rounds.

3. Socialization of unacquainted members of groups. Division of participants into two groups of 6 people.

4. Participants play the games (PD and UG or TG) in the newly formed groups for 15 rounds.

5. Participants take tests.

6. Participants are given money for the experiment.

The number of experiments involving PD and UG is 18; involving PD and TG, 19.

## 3. Results and Discussion

### 3.1. The Prisoner's Dilemma

First of all, let us consider the differences in cooperation rate before and after socialization for the Russians and for the Yakuts with a reference to gender.

Before socialization, the cooperation rate is significantly higher for the Yakuts ($Z = -5.894$, $p < 0.0001$, Wilcoxon rank-sum test) (Table 3). After socialization, the cooperation rate is higher for the Russians ($Z = 2.52$, $p = 0.01$, Wilcoxon rank-sum test). This means that socialization works better for the Russians.

**Table 3**

Cooperation rates before and after socialization for the Russians and for the Yakuts.

| Prisoner's dilemma | Cooperation before socialization | | Cooperation after socialization | | Z | p-value |
|---|---|---|---|---|---|---|
| | M | SD | M | SD | | |
| the Russians | 0.22 | 0.22 | 0.58 | 0.38 | -9.49 | <0.0001 |
| the Yakuts | 0.37 | 0.20 | 0.49 | 0.29 | -3.93 | 0.0001 |
| Z | -5.89 | | 2.52 | | | |
| p-value | <0.0001 | | 0.01 | | | |

Results are from Wilcoxon rank sum test (for independent measurements) and Wilcoxon signed rank sum test (for repeated measures). Nonparametric methods were used because the data are not normally distributed.

In fact, for the Russians, there is no difference in cooperation rates among males and females before socialization ($Z = 0.9$, $p = 0.37$, Wilcoxon rank-sum test); whereas for the Yakuts, before socialization females cooperate by 0.13 more often than males ($Z = 3.6$, $p = 0.003$, the Wilcoxon rank-sum test) (Table 4).

**Table 4**

Cooperation rates before and after socialization for the Russians' and the Yakuts' males and females.

| Prisoner's dilemma | Cooperation before socialization | | Cooperation after socialization | | Z | p-value |
|---|---|---|---|---|---|---|
| | M | SD | M | SD | | |
| the Russians' males | 0.21 | 0.21 | 0.61 | 0.41 | -7.85 | <0.0001 |
| the Russians' females | 0.24 | 0.22 | 0.54 | 0.43 | -5.28 | <0.0001 |
| Z | 0.9 | | -0.81 | | | |
| p-value | 0.37 | | 0.42 | | | |
| the Yakuts' males | 0.29 | 0.20 | 0.45 | 0.32 | -2.54 | 0.01 |
| the Yakuts' females | 0.42 | 0.19 | 0.52 | 0.27 | -3.01 | 0.003 |
| Z | 3.6 | | 1.47 | | | |
| p-value | 0.0003 | | 0.14 | | | |

Results are from Wilcoxon rank sum test (for independent measurements) and Wilcoxon signed rank sum test (for repeated measures). Nonparametric methods were used because the data are not normally distributed.

After socialization, there is no gender difference in cooperation rates for the Russians as well as for the Yakuts (Moscow: $Z = -0.81$, $p = 0.42$; Yakutsk: $Z = 1.47$, $p = 0.14$, Wilcoxon rank-sum test) (Table 4). Hence, socialization equalizes the gender differences in cooperation rates within groups.

After socialization, there a difference between the Russians' males and the Yakuts' males in cooperation rate ($Z = 2.67$, $p = 0.008$, Wilcoxon rank-sum test), but the Yakuts' and the Russians' females demonstrate similar behavior patterns in cooperation after socialization ($Z = 0.82$, $p = 0.41$, rank-sum test).

### 3.2. The Ultimatum Game

The general indicators of the UG are the number of points, which the first player offers to the second player and the second player's decision to accept or reject this offer. These are so called proposal rates and rejection rates.

In the framework of the UG, socialization does not reveal a statistically significant increase in proposal and rejections rates for the Yakuts, but for the Russians, these rates differ (Table 5).

**Table 5**

Proposal rates before and after socialization for the Russians and for the Yakuts.

| Ultimatum Game | Proposal before socialization | | Proposal after socialization | | Z | p-value |
|---|---|---|---|---|---|---|
| | M | SD | M | SD | | |
| the Russians | 4.00 | 0.60 | 4.56 | 0.56 | -9.43 | <0.0001 |
| the Yakuts | 4.32 | 0.68 | 4.85 | 0.82 | -1.18 | 0.24 |
| Z | -2.03 | | -0.98 | | | |
| p-value | 0.04 | | 0.33 | | | |

Results are from Wilcoxon rank sum test (for independent measurements) and Wilcoxon signed rank sum test (for repeated measures). Nonparametric methods were used because the data are not normally distributed.

Before socialization, proposal rate is higher for the Yakuts ($Z$ = -2.03, $p$ = 0.04, Wilcoxon rank-sum test) (Table 5). However, after socialization, there is no difference between the Russians and the Yakuts in terms of proposals ($Z$ = -0.98, $p$ = 0.33, Wilcoxon rank-sum test).

Moreover, there is no difference among these ethnicities in terms of rejection rates before socialization; but after socialization rejection rate for the Yakuts is higher than for the Russians ($Z$ = -3.15, $p$ = 0.002, Wilcoxon rank-sum test) (Table 6).

**Table 6**

Rejection rates before and after socialization for the Russians and for the Yakuts.

| Ultimatum Game | Reject % before socialization | | Reject % after socialization | | Z | p-value |
|---|---|---|---|---|---|---|
| | M | SD | M | SD | | |
| the Russians | 0.10 | 0.12 | 0.04 | 0.08 | 6.89 | <0.0001 |
| the Yakuts | 0.12 | 0.09 | 0.06 | 0.10 | 1.54 | 0.12 |
| Z | -1.07 | | -3.15 | | | |
| p-value | 0.28 | | 0.002 | | | |

Results are from Wilcoxon rank sum test (for independent measurements) and Wilcoxon signed rank sum test (for repeated measures). Nonparametric methods were used because the data are not normally distributed.

Before socialization, the Russians' females make more generous offers than the Russians' males ($Z$ = 2.25, $p$ = 0.02, Wilcoxon rank-sum test), but for the Yakuts, no gender difference is observed ($Z$ = 1.28, $p$ = 0.2, Wilcoxon rank-sum test) (Table 7). After socialization, the difference between males and females disappears for both the Russians and the Yakuts (the Russians: $Z = 0.39$, $p = 0.7$; the Yakuts: $Z = 0.49$, $p = 0.62$, Wilcoxon rank-sum test).

**Table 7**

Proposal rates before and after socialization for the Russians' and the Yakuts' males and females.

| Ultimatum Game | Proposal rate before socialization | | Proposal rate after socialization | | Z | p-value |
|---|---|---|---|---|---|---|
| | M | SD | M | SD | | |
| the Russians' males | 3.93 | 0.57 | 4.56 | 0.57 | -7.71 | <0.0001 |
| the Russians' females | 4.09 | 0.60 | 4.55 | 0.53 | -5.47 | <0.0001 |
| Z | 2.25 | | 0.39 | | | |
| p-value | 0.02 | | 0.7 | | | |
| the Yakuts' males | 4.18 | 0.59 | 4.68 | 0.71 | -1.15 | 0.25 |
| the Yakuts' females | 4.45 | 0.78 | 5.01 | 0.95 | -0.21 | 0.83 |
| Z | 1.28 | | 0.49 | | | |
| p-value | 0.2 | | 0.62 | | | |

Results are from Wilcoxon rank sum test (for independent measurements) and Wilcoxon signed rank sum test (for repeated measures). Nonparametric methods were used because the data are not normally distributed.

### 3.3. The Trust Game

The general indicators of the TG are the amount of points, which the first player offers to the second player and the amount of points, which the second player transfers to the first. These indicators can be termed as the trust rate (from the first player) and the trustworthiness rate (from the second player).

First of all, it was discovered that in this game socialization does not work. The behavior of males and females is similar before and after socialization (Table 8).

**Table 8**

Trust rates before and after socialization for the Russians and for the Yakuts.

| Trust game | Trust before socialization | | Trust after socialization | | Z | p-value |
|---|---|---|---|---|---|---|
| | M | SD | M | SD | | |
| the Russians | 4.36 | 3.40 | 8.23 | 2.43 | -5.8 | <0.0001 |
| the Yakuts | 4.32 | 2.53 | 4.71 | 3.17 | -0.6 | 0.6 |
| Z | -0.24 | | 6.3 | | | |
| p-value | 0.81 | | <0.0001 | | | |

Results are from Wilcoxon rank sum test (for independent measurements) and Wilcoxon signed rank sum test (for repeated measures). Nonparametric methods were used because the data are not normally distributed.

Given this circumstance, participants' behavior patterns for the Russians and for the Yakuts do not differ before socialization ($Z$ = -0.24, $p$ = 0.81, Wilcoxon rank-sum test). However, after socialization, the Russians participants' trust rates increased by a factor of two compared to that of the Yakuts participants' ($Z$ = 6.3, $p$ = <0.0001, Wilcoxon rank-sum test).

Regarding the rate of trustworthiness, we register a similar situation, i.e. before socialization, there is no distinction between the Russians' and the Yakuts' participants ($Z$ = 0.55, $p$ = 0.59, Wilcoxon rank-sum test), but after socialization, the trustworthiness rate is significantly higher for the Russians than for the Yakuts (Z = 5.93, p < 0.0001, Wilcoxon rank-sum test) (Table 9).

**Table 9**

Trustworthiness rates before and after socialization for the Russians and for the Yakuts.

| Trust game | Trustworthy before socialization | | Trustworthy after socialization | | Z | p-value |
|---|---|---|---|---|---|---|
| | M | SD | M | SD | | |
| the Russians | 3.27 | 2.44 | 9.32 | 5.24 | -5.52 | <0.0001 |
| the Yakuts | 3.03 | 2.23 | 3.81 | 4.00 | -0.62 | 0.54 |
| Z | 0.55 | | 5.93 | | | |
| p-value | 0.59 | | < 0.0001 | | | |

Results are from Wilcoxon rank sum test (for independent measurements) and Wilcoxon signed rank sum test (for repeated measures). Nonparametric methods were used because the data are not normally distributed.

There is no gender difference for both the Russians and the Yakuts before socialization (the Russians: $Z = 0.2$, $p = 0.84$, the Yakuts: $Z = 1.64$, $p = 0.1$, Wilcoxon rank-sum test). For the Yakuts, no gender difference is observed after socialization ($Z = -0.16$, $p = 0.87$, Wilcoxon rank-sum test). However, for the Russians, males tend to trust more than females after socialization ($Z = -2.44$, $p = 0.02$, Wilcoxon rank-sum test) (Table 10).

**Table 10**

Trust rates before and after socialization for the Russians' and the Yakuts' males and females.

| Trust game | Trust before socialization | | Trust after socialization | | Z | p-value |
|---|---|---|---|---|---|---|
| | M | SD | M | SD | | |
| the Russians' males | 4.40 | 3.59 | 8.77 | 1.97 | -5.209 | <0.0001 |
| the Russians' females | 4.24 | 2.88 | 6.59 | 2.97 | -2.589 | 0.01 |
| Z | 0.2 | | -2.44 | | | |
| p-value | 0.84 | | 0.02 | | | |
| the Yakuts' males | 3.23 | 2.43 | 9.70 | 5.32 | -1.43 | 0.15 |
| the Yakuts' females | 4.62 | 2.52 | 4.65 | 3.14 | 0.54 | 0.59 |

| | | | |
|---|---|---|---|
| Z | 1.64 | -0.16 | |
| p-value | 0.1 | 0.87 | |

Results are from Wilcoxon rank sum test (for independent measurements) and Wilcoxon signed rank sum test (for repeated measures). Nonparametric methods were used because the data are not normally distributed.

Before socialization, the Yakuts' females demonstrate reciprocal gratitude more often than males ($Z = 2.23$, $p = 0.03$, Wilcoxon rank-sum test). In the other cases, no gender difference is revealed (the Russians before socialization: $Z = 0.26$, $p = 0.79$; the Russians after socialization: $Z = -0.83$, $p = 0.41$; the Yakuts after socialization: $Z = -0.07$, $p = 0.94$, Wilcoxon rank-sum test) (Table 11).

**Table 11**

Trustworthiness rates before and after socialization for the Russians' and the Yakuts' males and females.

| Trust game | Trustworthiness before socialization | | Trustworthiness after socialization | | Z | p-value |
|---|---|---|---|---|---|---|
| | M | SD | M | SD | | |
| the Russians' males | 3.92 | 2.62 | 4.79 | 3.38 | -4.75 | <0.0001 |
| the Russians' females | 3.37 | 2.58 | 8.19 | 5.01 | -2.82 | 0.005 |
| Z | 0.26 | | -0.83 | | | |
| p-value | 0.79 | | 0.41 | | | |
| the Yakuts' males | 2.67 | 2.49 | 3.98 | 4.25 | -1.22 | 0.22 |
| the Yakuts' females | 3.29 | 2.01 | 3.68 | 3.91 | 0.3 | 0.76 |
| Z | 2.23 | | -0.07 | | | |
| p-value | 0.03 | | 0.94 | | | |

Results are from Wilcoxon rank sum test (for independent measurements) and Wilcoxon signed rank sum test (for repeated measures). Nonparametric methods were used because the data are not normally distributed.

Building a chart, we can visualize the dependences of the sum that the first player offers to the second player and the second player's degree of reciprocity, manifested in a sum that is transferred back to the first player. For example, if the first player offers five points, and the second player transfers eight points in reciprocity; we obtain a data point (5, 8) in our chart.

The coefficient of the slope of the line that extrapolates data points on the chart represents the degree of reciprocal gratitude. Here, it is important to note that, for both the Russians and the Yakuts, the line equation is y=0.75x-0.31 ($R^2 = 0.79$) before socialization (Fig. 1); but after socialization, the equation is y=1.25x-1.57 ($R^2 = 0.73$) (Fig. 2). Hence, it obvious that degree of reciprocal gratitude increases by a factor of two. Besides, before socialization, the coefficient of the slope of the line is less than one, after socialization it is already more than one in general and in each city. This illustrates an increase in the rate of trust among participants.

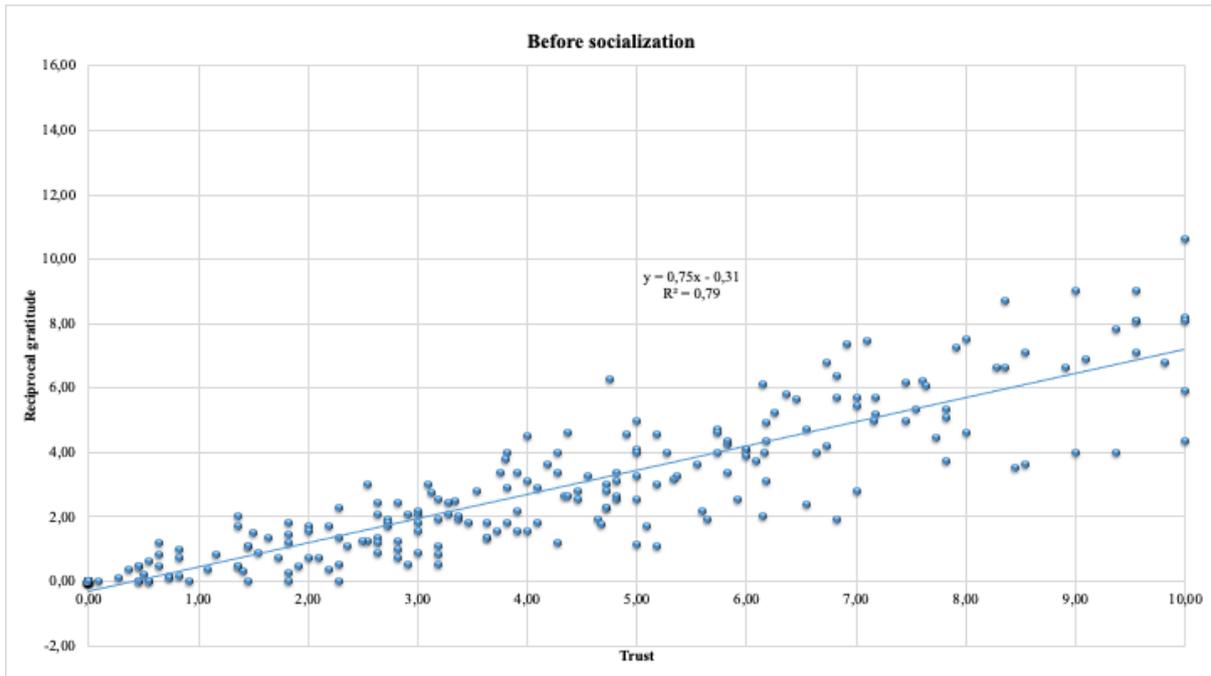

**Fig. 1.** The dependence of the sum that the first player offers to the second player and the second player's degree of reciprocity before socialization

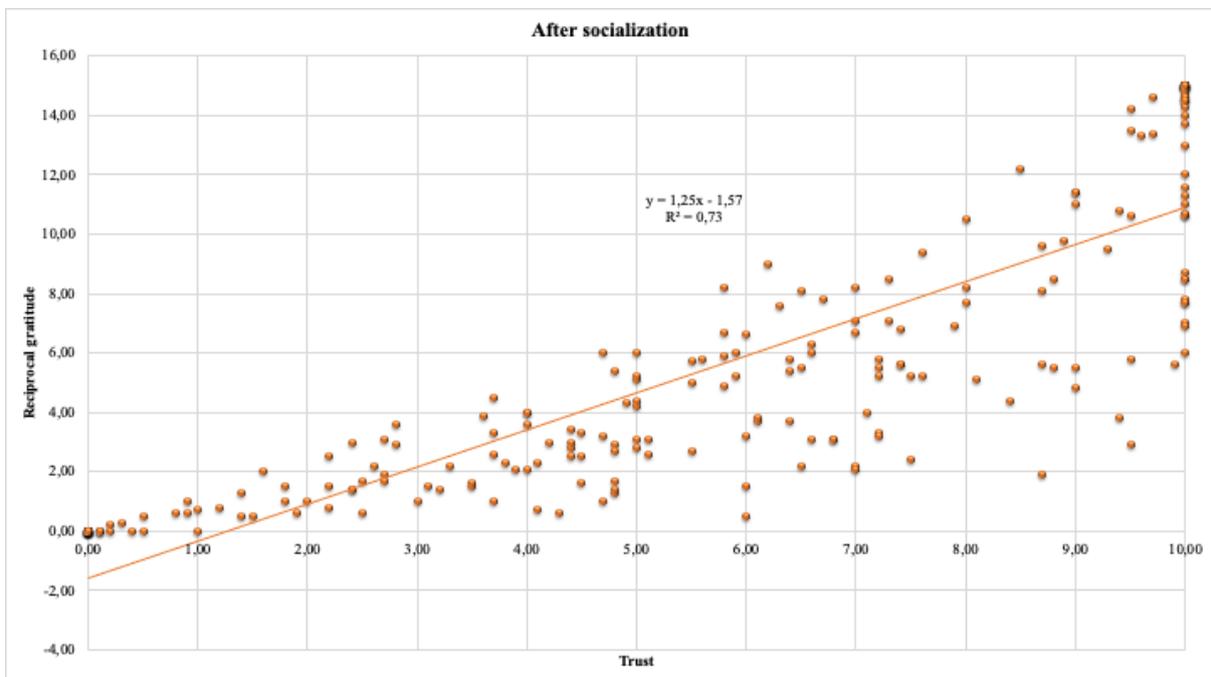

**Fig. 2.** The dependence of the sum that the first player offers to the second player and the second player's degree of reciprocity after socialization

For the Russians, the line of the dependence of the sum that the first player sends to his partner and the sum that the partner transfers back to the first player in reciprocity before socialization corresponds to the equation y = 0.88x – 0.58 ($R^2$ = 0.95) (Fig. 3), and, after socialization, the coefficient of the slope of the line increases by a factor of two: y = 1.66x – 4.3 ($R^2$ = 0,66) (Fig. 4).

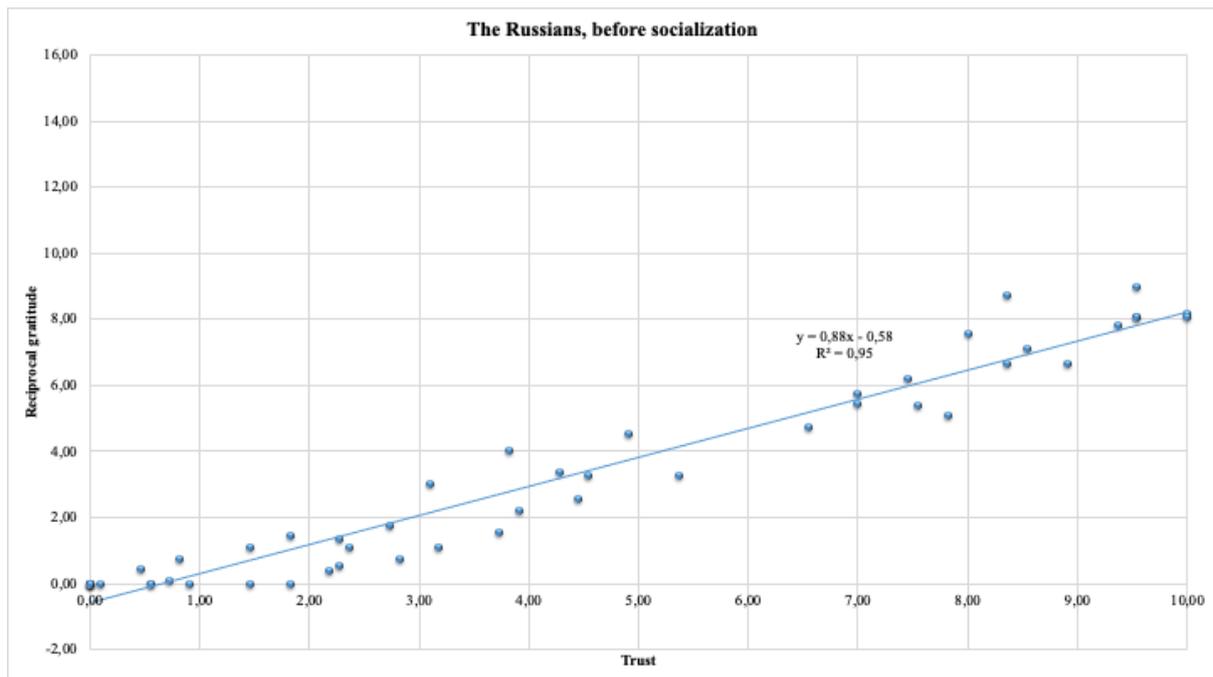

**Fig. 3.** The dependence of the sum that the first player offers to the second player and the second player's degree of reciprocity before socialization for the Russians.

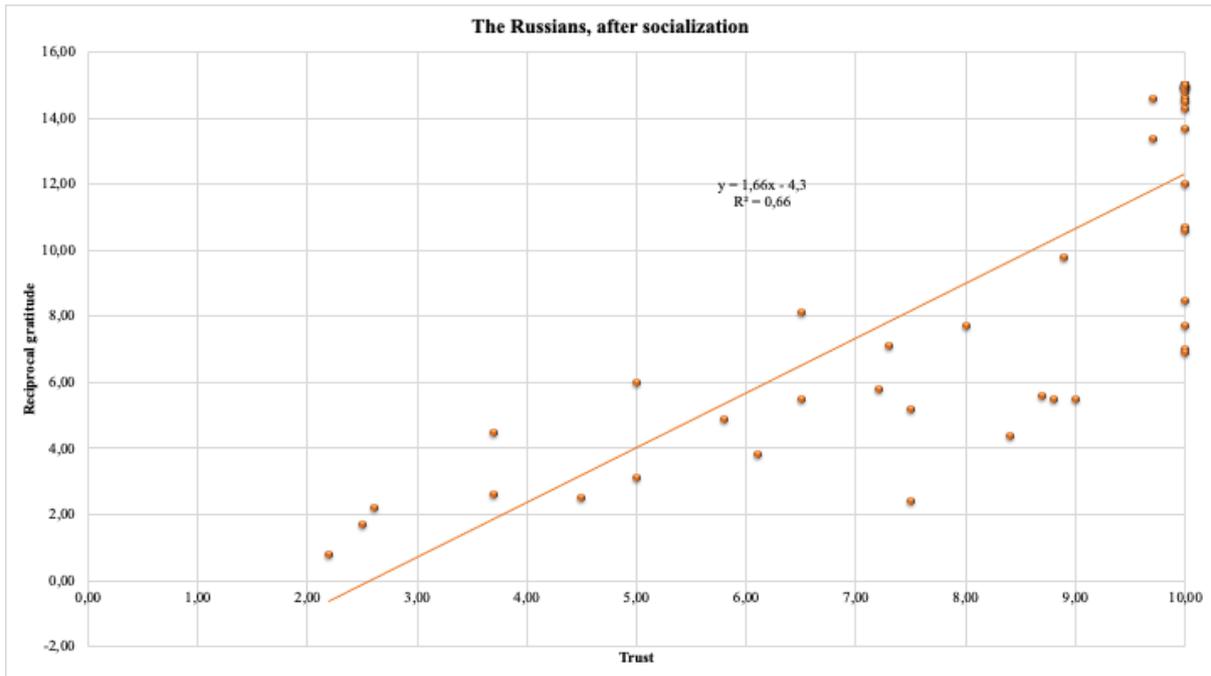

**Fig. 4.** The dependence of the sum that the first player offers to the second player and the second player's degree of reciprocity for the Russians.

For the Yakuts, the coefficient of the slope of the line also increases before socialization and can presented with the equation y = 0.71x – 0.13 ($R^2$ = 0.7) (Fig. 5). After socialization, the equation is y = 1.18x – 1.28 ($R^2$ = 0.71) (Fig. 6).

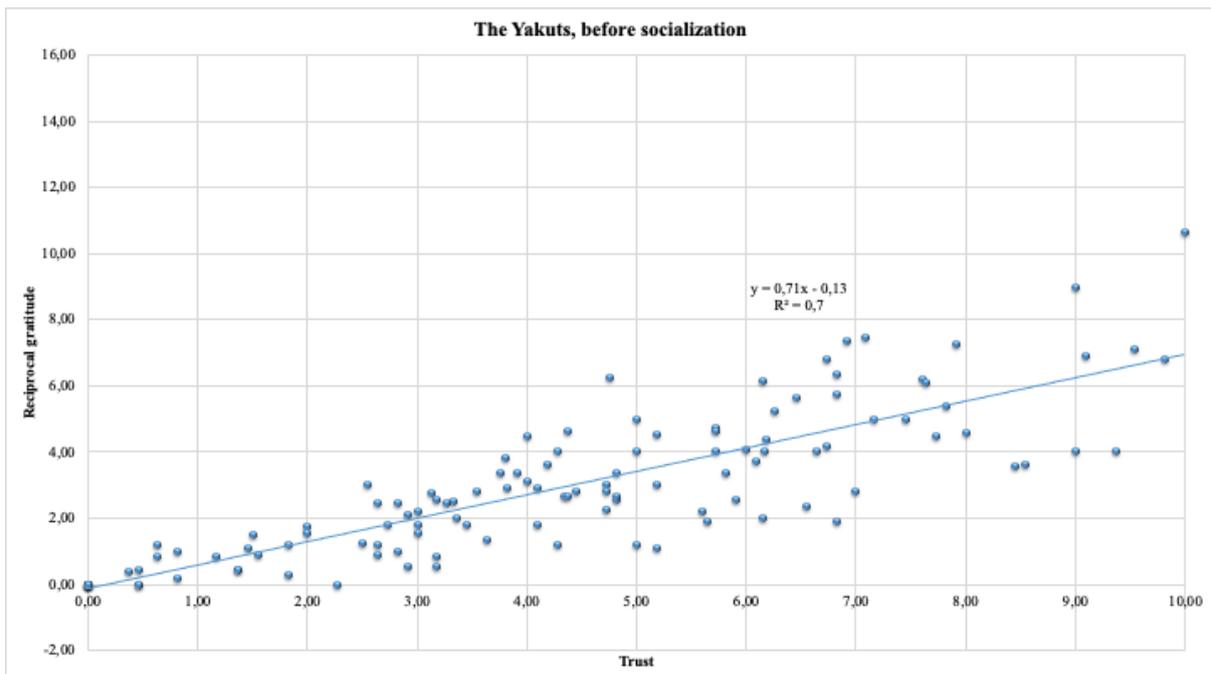

**Fig. 5.** The dependence of the sum that the first player offers to the second player and the second player's degree of reciprocity before socialization for the Yakuts.

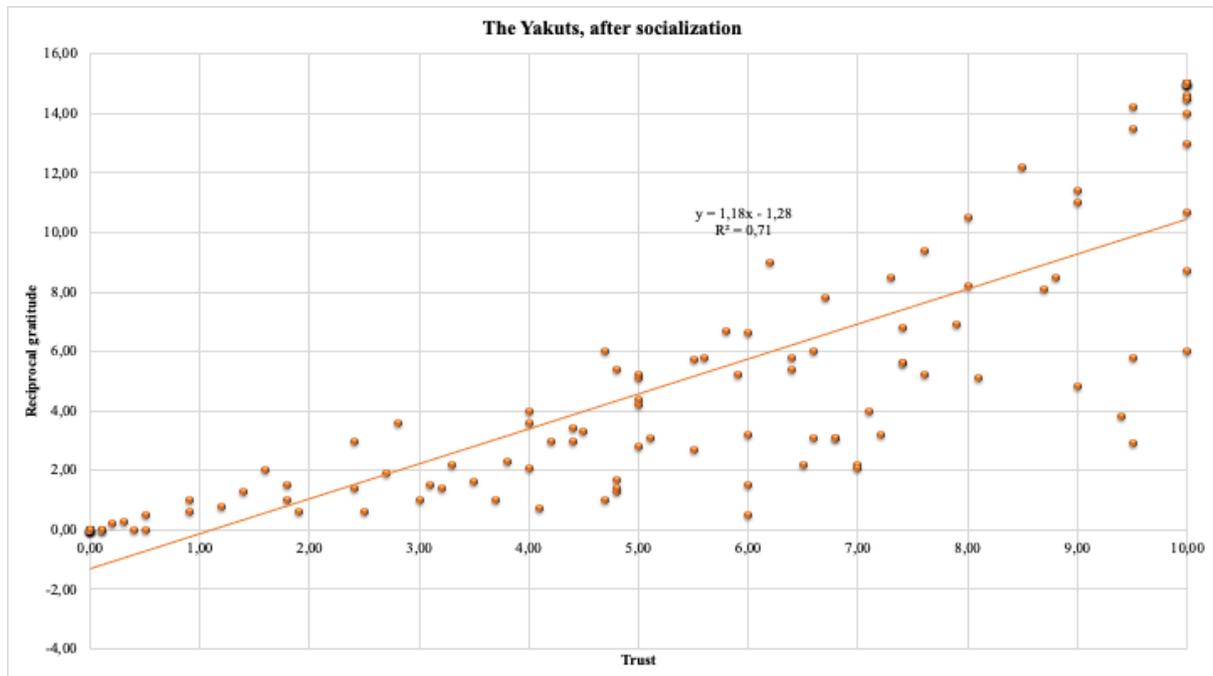

**Fig. 6.** The dependence of the sum that the first player offers to the second player and the second player's degree of reciprocity after socialization for the Yakuts.

The coefficients of the reciprocal behavior are similar before socialization for the Russians and the Yakuts. However, after socialization, the coefficient for the Russians is higher than for the Yakuts, which correlates with the previous results.

### 3.4. General patterns

The cooperation rates correlate with the trust rates before socialization (Spearman's *rho* = 0.5, *p* < 0.0001). This correlation is higher for the Yakuts (Spearman's *rho* = 0.74, *p* =0.006) than for the Russians (Spearman's *rho* = 0.47, *p* < 0.0001).

Besides, it was found that the rejection rate after cooperation correlates with the cooperation rate before and after socialization. The correlation between the cooperation rate before socialization and the rejection rate after socialization is equal to 0.75 ($p = 0.09$); between the cooperation rate and rejection rate after cooperation, 0.93 ($p = 0.01$).

In the TG, the cooperation rate after socialization correlates with the trust rate after socialization for the Yakuts (Spearman's rho = 0.51, $p < 0.0001$).

## 4. Conclusion

Thus, the study of the effect of ethnicity and gender differences on economic behavior in a multinational state was performed for the Russians (the indigenous and main population in Russian Federation) and for Yakuts (the indigenous nomadic population of eastern type in Russian Federation) using the Prisoner's Dilemma Game, Ultimatum Game, and Trust Game. In-group behavior was studied before socialization and out-group behavior, after socialization. This study is the continuation of social utility investigation (Babkina et al., 2016; Berkman et al., 2015). It was found that socialization function is a mechanism for increasing the rate of cooperation. However, this study showed that socialization does not perform this function in all social dilemmas and does not apply to all cultures. According to our findings, socialization increases the cooperation rate for the Russians most, compared to the Yakuts. The Russians participants started to cooperate and trust more in the Ultimatum Game and Trust Game after socialization, whereas the Yakuts participants' behavior after socialization was similar to their behavior before socialization.

In the Prisoner's Dilemma Game, ethnical differences are registered in participants' behavior before and after socialization. Before socialization, males and females differ in terms of their cooperation rate; but after socialization, the statistically significant difference disappears.

In the Ultimatum Game, there is a difference between the Russians and the Yakuts that consists in proposal rates before and after socialization. However, the rejection rate is the same for both populations. Males' behavior is similar to that of females.

In the Trust Game, the Russians and the Yakuts demonstrate the same rates of trust and trustworthiness among participants of the experiments with the sole exception of the trust rate before socialization. Before socialization, males and females do not differ in trust, but after socialization, males tend to trust more than females in both the cities. The reciprocal behavior is identical in males and females before and after socialization. However, after socialization, participants start to display more trust and express reciprocal in both the cities.

It was also concluded that the non-cooperative Prisoner's Dilemma game is perceived in a similar way both for the Russians and the Yakuts. However, the Yakuts show stable behavior patterns in bargaining games (Ultimatum game and Trust game), i.e. their behavior in groups with strangers is similar to the behavior in socialized groups. This can be due to the fact that even a non-socialized group is perceived by these participants as their "own", since it comprises compatriots. Finally, no gender difference manifests itself in bargaining games.

## Acknowledgments

We thank Rinat Yaminov for writing the programing code for experiments, Alexander Chaban for technical help in conducting experiments at MIPT. We are grateful for comments from Ivan Kozitsin and anonymous reviewers. This research was supported by The Tomsk State University competitiveness improvement program and Russian Foundation for Basic Research 19-01-00296A.

**Web references**